\documentclass[twocolumn,showpacs,prl]{revtex4}
\usepackage{amssymb}
\usepackage{amsmath}
\usepackage{graphicx}
\usepackage{bm}
\usepackage{verbatim}

\usepackage{graphicx}
\usepackage{bm}

\begin{document}

\title{Vortex Core States in Superconducting Graphene.}
\author{ I.~M.~Khaymovich$^{(1)}$, N.~B.~Kopnin $^{(2,3)}$, A.~S.~Mel'nikov $^{(1)}$,
I.~A.~Shereshevskii $^{(1)}$,}
\affiliation{$^{(1)}$Institute for Physics of Microstructures, Russian Academy of
Sciences, 603950 Nizhny Novgorod, GSP-105, Russia}
\affiliation{$^{(2)}$ Low Temperature Laboratory, Helsinki University of
Technology, P.O. Box 2200, FIN-02015 HUT, Finland}
\affiliation{$^{(3)}$ L.~D.~Landau Institute for Theoretical Physics, 117940 Moscow,
Russia}

\date{\today}

\begin{abstract}
The distinctive features of the electronic structure of vortex
states in superconducting graphene are studied within the
Bogolubov--de Gennes theory applied to excitations near the Dirac
point. We suggest a scenario describing the subgap spectrum
transformation which occurs with a change in the doping level. For
an arbitrary vorticity and doping level we investigate the problem of
existence of zero energy modes. The crossover to a Caroli -- de
Gennes -- Matricon type of spectrum is studied.
\end{abstract}

\maketitle

Recent exciting developments in transport experiments on graphene
\cite{Novoselov05} have stimulated theoretical studies of possible
superconductivity phenomena in this material
\cite{Doniach07,CastroNeto05,Honerkamp,KopninSonin08}. A number of
unusual features of superconducting state have been predicted,
which are closely related to the Dirac-like spectrum of normal
state excitations. In particular, the unconventional normal
electron dispersion has been shown to result in a nontrivial
modification of Andreev reflection \cite{been2} and Andreev bound
states in Josephson junctions \cite{been3}. In this Letter we
consider another generic problem illustrating the new physics of
Andreev scattering processes in graphene, namely, the electronic
spectrum in the core of a vortex that can presumably appear in the
superconducting graphene in presence of magnetic field. The
electronic vortex structure for Dirac fermions has been previously
studied in particle physics for a situation equivalent to the zero
doping limit in graphene, when the Dirac point lies exactly at the
Fermi level. A number of important results have been obtained,
namely, the exact solutions for the zero energy modes
\cite{JackiwRossi81,Wilchek} and the subgap spectrum for some
model gap profiles within the vortex core \cite{Seradjeh08}. The
problem of zero energy modes has been further addressed in
\cite{seradjeh-franz,bergman} for vortices in various condensate
 phases described by the Dirac theory on a honeycomb
lattice.  Our goal is to develop a theoretical description of the
electronic structure of multiply quantized vortices beyond the
zero doping limit and study the transformation of the spectrum
under the shift $\mu$ of the Fermi level from the Dirac point.

The energy spectrum is determined by the vortex winding number
(vorticity) $n$ and labeled by the angular momentum $\nu$ which is a conserved quantity
 for an axisymmetric vortex.
For zero doping the sub-gap spectrum consists of
$n$ zero energy states \cite{JackiwRossi81,Wilchek}.
 The states with higher
energies lie close to the gap edge $\pm |\Delta_0|$. In the
present Letter we demonstrate that, with an increase in doping
level $\mu$, the distance between the energy levels decreases, so
that more and more states fill the sub-gap region. The set of
low-energy levels gradually transforms into a set of $n$
``spectrum branches'' $E^{(i)}(\nu)$ where $i=1,\ldots , n$. If
$\nu$ is considered as a continuous parameter, these $n$ energy
branches cross the Fermi level \cite{volovik} as functions of
$\nu$. The detailed behavior of the energy spectrum as a function
of $\mu$ crucially depends on the parity of the winding number.
For {\em odd} $n$ there exists one branch which intersects zero
energy at $\nu =0$. Its crossing point belongs to the spectrum,
thus resulting in an exact zero energy mode. The crossing points
of other $n-1$ branches do not generally belong to the spectrum
for finite $\mu$, thus these zero modes do not exist. However,
some of these $n-1$ zero energy modes can appear again for certain
doping levels. For high doping $|\mu|\gg |\Delta_0|$, they appear
almost periodically with an increase in the Fermi momentum $\hbar
k_F=|\mu|/ v_F$ by a characteristic value $\delta k_F\sim 1/\xi$.
Here $\xi$ is a superconducting coherence length $\xi=\hbar
v_F/\Delta_0$, and $\Delta_0$ is a homogeneous gap value far from
the vortex center. In a singly quantized vortex the energy
spectrum for high doping is given by the Caroli--de
Gennes--Matricon (CdGM) expression similar to that for usual
s-wave superconductors \cite{CdGM}.  For an {\em even} winding
number the crossing points of all $n$ branches do not generally
belong to the spectrum if $\mu\neq 0$. In this case the exact zero
modes are absent. Nevertheless, some of these zero energy modes
can appear again for certain $\mu$ in the way similar to the odd
vorticity case. For high doping, they appear periodically with
increasing $\mu$. Indeed, the energy branches cross the Fermi
level at certain momenta $\nu^{(i)} \sim k_F\xi$. The zero modes
appear when these $\nu^{(i)}$ become integer (odd $n$) or
half-integer (even $n$) for discrete equally spaced (due to
equidistant energy levels) values of the Fermi momentum. Because
of the symmetry with respect to the sign of energy (see below) the
zero energy modes appear and disappear in pairs, for $\nu$ and
$-\nu$.

\paragraph{Model.}
The Bogolubov--de Gennes (BdG) equations in superconducting
graphene can be written as two decoupled sets of four equations
each \cite{been2}. Assuming valley degeneracy we can consider only
one of these sets known as "Dirac--BdG equations"
\cite{castro-neto05,been2}
\begin{eqnarray}
v_F\hat {\bm \sigma}\cdot \left(\check {\bf p} -\frac{e}{c}{\bf
A}\right) \hat u +\Delta \hat v =(E+\mu)\hat u \ , \label{DirBdG-u} \\
-v_F\hat {\bm \sigma}\cdot \left(\check{\bf p}+\frac{e}{c}{\bf
A}\right) \hat v +\Delta^* \hat u =(E-\mu)\hat v \ .
\label{DirBdG-v}
\end{eqnarray}
Here $\check {\bf p}= -i\hbar {\bm \nabla}$, and $ \hat u =(u_1, \
u_2 )$ and $\hat v =(v_1, \ v_2) $ are two-component wave functions
of electrons and holes in different valleys in the Brillouin zone
\cite{CastroNeto08}.
The indices 1 and 2 denote two sublattices of the honeycomb structure.
 They form spinors in the pseudo-spin space,
in which the Pauli matrices $ \hat \sigma _x$, $\hat \sigma_y$,
$\hat \sigma _z$ are defined; we also introduce a
 vector $\hat {\bm \sigma}=(\hat \sigma _x , \; \hat\sigma_y)$. The energy is
measured from the Fermi level. For zero doping, the Fermi level
lies exactly at the Dirac point, $\mu =0$. For nonzero doping, the
Fermi level is shifted by $\mu>0$ upwards (electron doping) or
downwards $\mu <0$ (hole doping) from the Dirac point.
For a homogeneous gap $\Delta =\Delta_0$ and zero
 magnetic field  the
wave functions take the form of plane waves $ \hat u, \hat v
\propto e^{i{\bf p}\cdot {\bf r}}$ and
 we obtain two noninteracting
energy branches $E^2 = \Delta_0^2+(\mu\mp v_F p)^2$, which
correspond to the pseudospin orientation chosen parallel and antiparallel to the
momentum direction, respectively.

\paragraph{Vortex states.}
Consider an $n$-quantum vortex, $ \Delta =|\Delta(\rho)|e^{in\phi}
$, where $\rho, \phi$ are cylindrical coordinates. For an
axisymmetric magnetic field the eigenstates are labelled by a
discrete angular momentum quantum number $\nu$:
\begin{equation}
\left(\begin{array}{c}\hat u\\ \hat v
\end{array}\right)= e^{i\nu\phi-i\hat \sigma_z\phi /2+i\hat
\sigma_z\pi/4}
\left(\begin{array}{c}e^{in\phi /2} \hat U(\rho) \\
e^{-in\phi /2} \hat V(\rho)
\end{array}\right) \ . \label{transform}
\end{equation}
The extra factor $e^{-i\hat \sigma_z \phi /2}$ as compared to the
usual BdG functions $u$ and $v$ comes from the angular dependence
of the momentum operator in cylindrical coordinates $\check p_x
\pm i \check p_y = e^{\pm i\phi} \hbar [-i \partial /\partial \rho
 \pm \rho^{-1} \partial /\partial \phi]$.
The transformation properties of the wave functions under rotation
around the vortex axis correspond to those for an $s$-wave
superconductor with the replacement $n\rightarrow n- 1$. As a
result, $\nu$ is half-integer for even vorticity $n$ and integer
for odd $n$ in contrast to standard $s$-wave superconductors.

\paragraph{General properties of zero energy states.}
Equations (\ref{DirBdG-u}) and (\ref{DirBdG-v}) are invariant
under the transformation
\begin{equation}
E \rightarrow -E\ ,\; i\hat \sigma_y \hat u^*\rightarrow \hat v \
, \; i\hat \sigma_y \hat v^*\rightarrow - \hat u \ . \label{E-sym}
\end{equation}
Thus, for arbitrary vorticity and doping level, a set of zero
energy modes  $(\hat u_i, \hat v_i)$ labelled by an index $i$,
satisfies \cite{JackiwRossi81} $ \hat v_i=i\hat\sigma_y \hat u_j^*
$ , $ \hat u_i=-i\hat \sigma_y \hat v_j^*  $. This transformation
couples the states with opposite angular momenta. One can separate
two types of zero energy solutions: (i)  The eigenfunction
components transform into each other, $i=j$, which can be realized
only for $\nu =0$. (ii) There exists a pair of eigenfunctions
$i\neq j$ with opposite $\nu \ne 0$ coupled by the above
transformation.

Consider solutions of the first type which can exist only for an
odd-quantum vortex. We put $\hat v=i\hat \sigma_y \hat u^* $ and
find:
\begin{equation}
\left[ v_F\hat {\bm \sigma}\cdot \left(\check {\bf p}-
\frac{e}{c}{\bf A}
\right)- \mu \right] \hat
u + i\Delta \hat \sigma_y \hat u^*   =0 \ . \label{eq-zeroE}
\end{equation}
We look for the solution in the form $\hat u=\zeta (\rho)\hat
U^{(0)}$, where $\hat U^{(0)}$ is a normal state solution
of Eq.~(\ref{eq-zeroE}) with $\Delta=0$
%
%
and $\zeta $ is a real function satisfying
\[
\hat \sigma_x e^{i\hat \sigma _z\phi} \hat U^{(0)} \hbar v_F
\left(d \zeta /d\rho\right) =\hat\sigma_y \hat U^{(0)*}\Delta
\zeta
\]
In a homogeneous magnetic field $H$
the functions
$\hat U^{(0)}$
 diverge at large distances $\rho $ of the order
of the magnetic length $L_H=\sqrt{\hbar c/eH}$ except for a
discrete set of $\mu$ which correspond to the Landau energy
levels. However, similarly to the case of usual superconductors,
this large-distance divergence is cut-off either at the magnetic
field screening length (isolated vortex) or at the intervortex
distance $\sim L_H$ (flux lattice). Thus, we can disregard the
above mentioned divergence for the analysis of the states
localized near the vortex core at distances of the order of the
coherence length $\xi$. Considering weak magnetic fields near the
vortex core $H\ll H_{c2}$ such that $L_H\gg\xi=\hbar v_F/\Delta_0$
we can even neglect the vector potential at all. In this way we
obtain for $\mu>0$:
\[
\hat U^{(0)} = C\left( e^{-i\pi/4} e^{i(n-1)\phi/2} J_{(n-1)/2}
(k_F\rho) \atop e^{i\pi/4} e^{i(n+1)\phi/2} J_{(n+1)/2} (k_F\rho)
\right)\ ,
\]
\begin{equation}
\zeta(\rho) = e^{-K(\rho)}\ ; \; K(\rho)=\frac{1}{\hbar
v_F}\int_0^\rho |\Delta(\rho^\prime)|\, d\rho^\prime  \ .
\label{zeta-K}
\end{equation}

To address the problem of the zero energy solutions of the second
type for $\nu \ne 0$ we note that the symmetry transformation Eq.
(\ref{E-sym}) implies that under the transformation $E\rightarrow
-E $ and $\nu \rightarrow -\nu$ the functions $\hat U$ and $\hat
V$ in Eq. (\ref{transform}) change according to $V_2 \rightarrow
U_1$ and $V_1\rightarrow -U_2$. Therefore, for $E=0$, the
functions with opposite momenta coupled by Eq. (\ref{E-sym}) obey
$ V_{2,\nu}=U_{1,-\nu}$, $V_{1, \nu}=-U_{2,-\nu} $. Equations for
$ U_1$ and $U_2$ take the form:
\begin{eqnarray}
\left(\frac{d}{d\rho} -\frac{N_-}{\rho} +\frac{eA_\phi}{\hbar
c}\right)U_{1,\nu} +\frac{|\Delta|}{\hbar v_F} U_{1,-\nu}\!\!
&=&\!\! \frac{\mu}{\hbar v_F} U_{2,\nu} \ , \quad \label{eq-XY1}\\
\left(\frac{d}{d\rho}+\frac{N_+}{\rho} -\frac{eA_\phi}{\hbar
c}\right)U_{2,\nu} +\frac{|\Delta|}{\hbar v_F} U_{2,-\nu}\!\! &=&
\!\! -\frac{\mu}{\hbar v_F} U_{1,\nu}\ .\qquad \label{eq-XY2}
\end{eqnarray}
Here $N_\pm = \nu +(n\pm 1)/2$ and $A_\phi(\rho)$ is the $\phi$-
component of the vector potential. We multiply the first equation
with $U_{2,\nu}$ and the second one with $U_{1,\nu}$. Next we do
the same for the opposite sign of $\nu$ and then add all the
resulting equations together. We thus find:
\begin{eqnarray}
&&\mu \int _0^\infty \left(U_{2,\nu}^2-U_{2,-\nu}^2-U_{1,\nu}^2
+U_{1,-\nu}^2
\right)\, \rho\, d\rho \nonumber \\
&=&\hbar v_F \int_0^\infty \frac{d}{d\rho}\left[ \rho ( U_{1,\nu}
U_{2,\nu}-U_{1,-\nu} U_{2,-\nu})\right]\, d\rho\   .\;
\label{condition-gen}
\end{eqnarray}
For a solution regular at the origin and decaying at $\rho
\rightarrow \infty$ the r.h.s. of Eq. (\ref{condition-gen}) is zero and,
thus, the l.h.s. of this equation also should
vanish. One can see that for $\nu \ne 0$ and $\mu \ne 0$, the
l.h.s. is nonzero in general. Therefore, zero-energy levels do not
generally exist.

To illustrate this we analyze the case of small $\mu$ using the
perturbation theory. For $\mu =0$ the equation for $U_1$ decouples
from that for $U_2$. As follows from \cite{JackiwRossi81} for
positive circulation $n\geq 1$, the functions $U_{1,\nu}$ are
regular while $U_{2,\nu}=0$ if $-\frac{1}{2}(n-1)\leq \nu \leq
\frac{1}{2}(n-1) $. For negative circulation, the functions
$U_{2,\nu}$ are regular while $U_{1,\nu}=0$  if
$\frac{1}{2}(n+1)\leq \nu \leq -\frac{1}{2}(n+1)$. In total there
exist $|n|$ zero energy levels. For small $\mu$ we can calculate
the integral in the l.h.s. of Eq. (\ref{condition-gen}) using the
wave functions satisfying Eqs. (\ref{eq-XY1}) - (\ref{eq-XY2})
with $\mu =0$. Using the results of Ref. \cite{JackiwRossi81} for
$n>0$, one can check that $ U_{\nu}^2-U_{-\nu}^2$ is nonzero
unless $\nu =0$. Therefore, there is no decaying solution and,
thus, there is no zero energy level for small $\mu$ and $\nu \ne
0$. However, the zero-energy levels may appear occasionally for
certain finite values of $\mu$, as it is shown below.

\paragraph{Large doping levels. Quasiclassical equations.}
The limit of large $\mu$ is very instructive and helps to get the
complete picture of the spectrum transformation. For the analysis
we follow a standard quasiclassical scheme (see \cite{MRS} for
details) and introduce a momentum representation: $ \psi ({\bm
\rho})  = (2\pi\hbar)^{-2}\int d^2p e^{i{\bf p}{\bm
\rho}/\hbar}\psi({\bf p})$, where ${\bf p} = p(\cos\theta_p,
\sin\theta_p)= p {\bf p}_0$. Let us transform the wave functions
so to choose the spin quantization axis along the direction of the
momentum ${\bf p}$: $\hat u = \hat S \hat g_u$, $\hat v = \hat S
\hat g_v$, where $\hat S =
e^{-i\theta_p\hat\sigma_z/2}(\hat\sigma_z+\hat\sigma_x)/\sqrt{2}$
and $\hat S^+
=(\hat\sigma_z+\hat\sigma_x)e^{i\theta_p\hat\sigma_z/2}/\sqrt{2}$.
Equations (\ref{DirBdG-u}), (\ref{DirBdG-v}) take the form:
\begin{eqnarray}
\left[ v_F \hat\sigma_z p -\mu \right]\hat g_u+ \hat h_{\check
A}\hat g_u+ \hat h_{\check \Delta}\hat g_v
 =E \hat g_u \ , \label{eq-p-1}\\
\left[ \mu- v_F \hat\sigma_z p\right]\hat g_v +\hat h_{\check
A}\hat g_v+ \hat h_{\check\Delta}^+ \hat g_u =E\hat g_v \ ,
\label{eq-p-2}
\end{eqnarray}
where we define the operators $\hat h_{\check A}= -(ev_F/c)\hat
S^+\hat {\bm \sigma} \check {\bf A}\hat S$, as well as $\hat
h_{\check \Delta}= \hat S^+\check \Delta\hat S$ and $\hat
h_{\check\Delta}^+= \hat S^+\check \Delta^*\hat S$. Here
 $\check{\bf A}={\bf A}(\check {\bm \rho})$
and $\check \Delta =\Delta(\check {\bm \rho})$ are functions of
the coordinate operator $\check {\bm \rho}=i\hbar
\partial/\partial {\bf p}$. Generally, the operators $\hat h_{\check A}$
and  $\hat h_{\check \Delta}$ mix the energy bands with opposite
pseudospin orientations with respect to the momentum direction.
However, taking the limit of large positive chemical potential
$\mu\gg\Delta_0$ and considering only the subgap spectrum one can
adopt the single band approximation with a fixed pseudospin
orientation:  $\hat\sigma_z \hat g_u =\hat g_u = g_u (1,0)$,
$\hat\sigma_z \hat g_v =\hat g_v= g_v (1,0)$. The accuracy of such
approximation can be determined using the second order
perturbation theory; the corresponding corrections to the energy
caused by the off-diagonal pseudospin terms in $\hat h_{\check A}$
and  $\hat h_{\check \Delta}$ are $ \delta E_A \sim (\xi^4/L_H^4)
(\Delta_0/k_F\xi) $, $\delta E_\Delta \sim \Delta_0/(k_F\xi)^3$.
These corrections are much smaller that the proper energy scale $
\Delta_0/k_F\xi$ for the sub-gap spectrum. One concludes therefore
that the single-band approximation is sufficient for the case of
large doping $\mu \gg \Delta$. The assumption of large $\mu>0 $
allows us to consider only the momenta close to $\hbar k_F$. Thus,
we put $p=\hbar k_F+q$ ($|q|\ll \hbar k_F$) and perform a Fourier
transformation into the $(s,\theta_p)$ representation:
\begin{equation}
\bar g({\bf p})=\left(g_u\atop g_v\right) = k_F^{-1}
\int_{-\infty}^{+\infty} ds e^{-iqs/\hbar} \bar g(s,\theta_p)\ .
\end{equation}
We introduce vector $\bar g$ and  Pauli matrices $\bar\tau_x,
\bar\tau_y, \bar\tau_z$ in electron-hole space. The angular
dependence can be separated: $\bar g(s,\theta_p) = \exp
(i\nu\theta_p+i\bar\tau_zn\theta_p/2)\bar F (s)$.

We start with an odd-quantum vortex and consider the low-energy
levels of the first kind which belong to the anomalous energy
branch crossing the Fermi level at $\nu =0$. For angular momenta
$\nu /k_F\xi \ll 1$, the solution can be found using a linear
expansion of $\hat h_{\check A}$ and $\hat h_{\check \Delta}$ in
terms of the angular momentum operator $\check\nu =
-i\partial/\partial\theta_p$. Assuming a small homogeneous
magnetic field, we obtain
\begin{eqnarray}
-i\hbar v_F\bar \tau_z\frac{\partial \bar F }{\partial s}  +
|\Delta (s)| \left[\frac{s}{|s|} \bar\tau_x +\frac{\nu n }{|s|k_F
}\bar\tau_y \right]\bar F = E^\prime \bar F \ . \label{andreev1}
\end{eqnarray}
where $E^\prime=E +\nu \hbar \omega_c/2 $ and
$\omega_c=eHv_F/c\hbar k_F$ is the cyclotron frequency; $\nu$ is
an integer.  Considering the last term in the above Hamiltonian as
a perturbation we find
\begin{equation}
\label{energy} E =-\nu\left[ \frac{n}{ Ik_F } \int_{0}^{\infty}
 \frac{|\Delta (s )|}{s}\,e^{-2K(s)}
 d s +\frac{\hbar \omega_c}{2}\right] \ ,
 \end{equation}
where $ I= \int_{0}^{\infty} \zeta ^2(s) \, d s$ while $K(s)$ and
$\zeta(s)$ are defined by Eq. (\ref{zeta-K}). Eq. (\ref{energy})
corresponds to the CdGM result \cite{CdGM} with an account of a
magnetic field in the spirit of Ref.~\cite{BrunHansen}. Since
$\nu$ can now take zero value, the spectrum has a zero energy
mode.

The levels of both first and second kind for any vorticity $n$ can
be easily described in the limit of large $\nu$. Here we can
replace the angular momentum operator by a classical continuous
variable. Instead of Eq. (\ref{andreev1}) we get from Eqs.
(\ref{eq-p-1}), (\ref{eq-p-2}) the Andreev equations along the
rectilinear quasiclassical trajectories:
\begin{equation}
\label{andreev} \left[-i\hbar v_F\bar
\tau_z\frac{\partial}{\partial s}+\frac{\hbar \omega_c k_F b}{2} +
\bar D({\bm \rho})-E
 \right] \bar g(s,\theta_p) = 0  ,
\end{equation}
where $\bar D = \bar\tau_x {\rm Re}\Delta ({\bm \rho}) -\bar\tau_y
{\rm Im}\Delta ({\bm \rho})$,  $b=-\nu/k_F$ is the trajectory
impact parameter, the position vector on the trajectory is ${\bm
\rho} =(s\cos\theta_p-b\sin\theta_p,s\sin\theta_p+b\cos\theta_p)$.
Eq. (\ref{andreev}) describes the quasiparticle states in an
arbitrary gap profile. For the particular case of a multiquantum
vortex the gap operator in Eq.(\ref{andreev}) takes the form: $
\Delta({\bm \rho}) = |\Delta (\sqrt{s^2+b^2})|
[(s+ib)/\sqrt{s^2+b^2}]^{n} e^{in\theta_p} $. For $|E|<\Delta_0$
the solution of corresponding Andreev equations is known to give
$|n|$ anomalous energy branches $E^{(i)}(b)$
 crossing zero energy as functions of a continuous parameter
 $b$ (see \cite{volovik,MRS} and references therein).

The total wave function should be single-valued. The appropriate
Bohr--Sommerfeld quantization rule for the angular momentum reads:
$k_F\int_0^{2\pi}b(\theta_p)d\theta_p = 2\pi m+ \pi(n-1)$, where
$m$ is an integer. The last term accounts for all the phase
factors that appear in $\hat S$ and $\hat g$. As a result, the
angular momentum $\nu$ should be integer or half-integer for odd
and even vorticity, respectively. This agrees with the conditions
imposed by Eq. (\ref{transform}).  The expression for the
anomalous branches at low energies take the form $E^{(i)}(b) \sim
\Delta_0(b-b_i)/\xi$, where $-\xi\lesssim b_i\lesssim \xi$. For
the branches with $b_i\neq 0$ the change in the Fermi momentum is
accompanied by the flow of the eigenvalues through the Fermi
level: the energy levels cross it by pairs at discrete values $k_F
=|\nu/b_i|$.

\paragraph{Numerical results.}
The above analytical considerations are in excellent agreement
with the results of our numerical calculations. For the numerical
solution, Eqs.~(\ref{DirBdG-u}), (\ref{DirBdG-v}) are expanded in
the basis of eigenfunctions for a normal graphene disk of radius
$R$. The vortex is placed at the disk center. We are not
interested here in the effect of boundaries on the subgap spectrum
and, thus, for the sake of simplicity we use the conditions of
zero current through the boundary in the form $u_1(R)=v_1(R) =0$.
To suppress the influence of the boundary condition the radius $R$
is taken much larger than the eigenfunction decay length in the
superconducting phase, i.e., the coherence length $\xi$.
\begin{figure}[t]
\centerline{\includegraphics[width=0.5\linewidth]{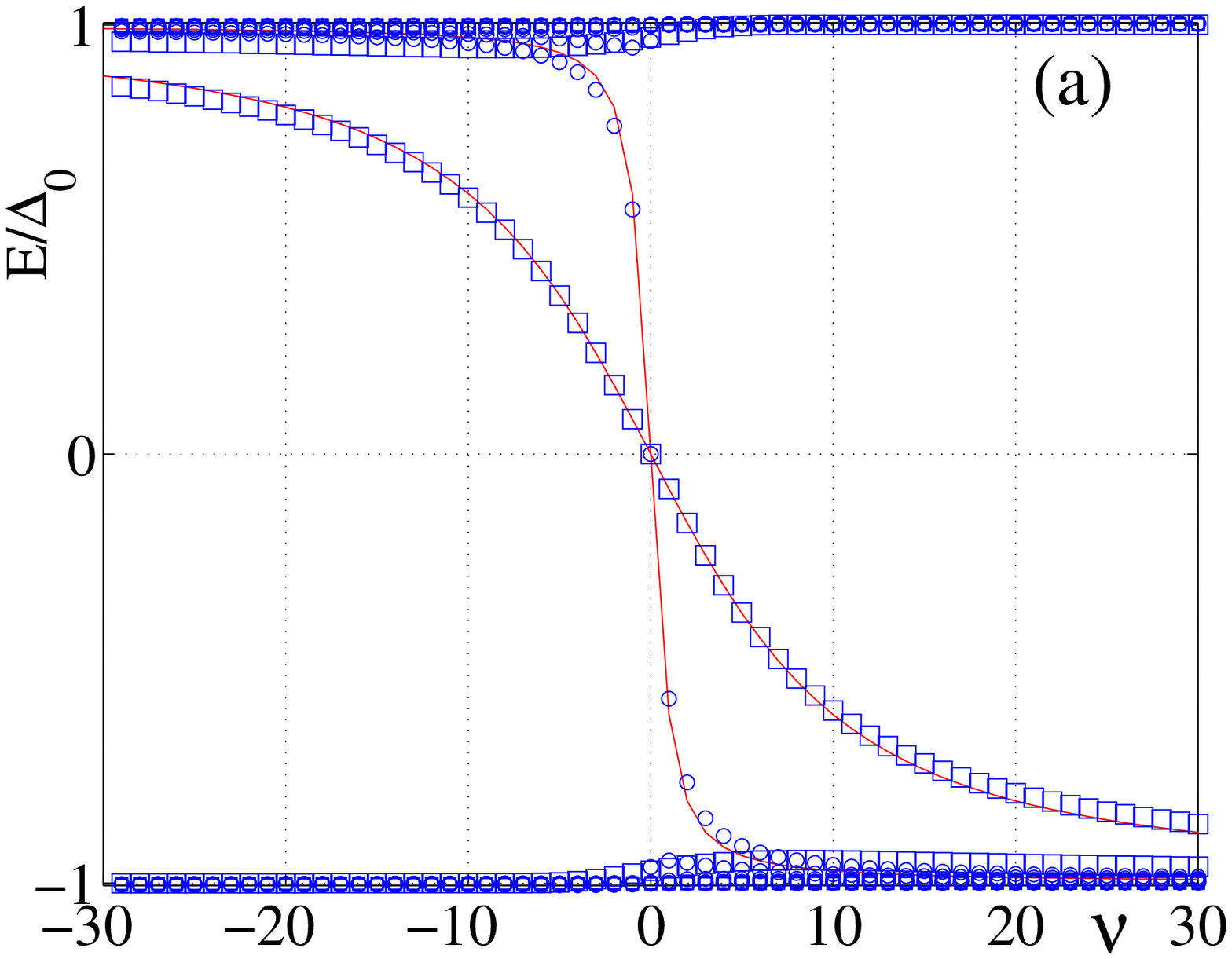}
\includegraphics[width=0.5 \linewidth]{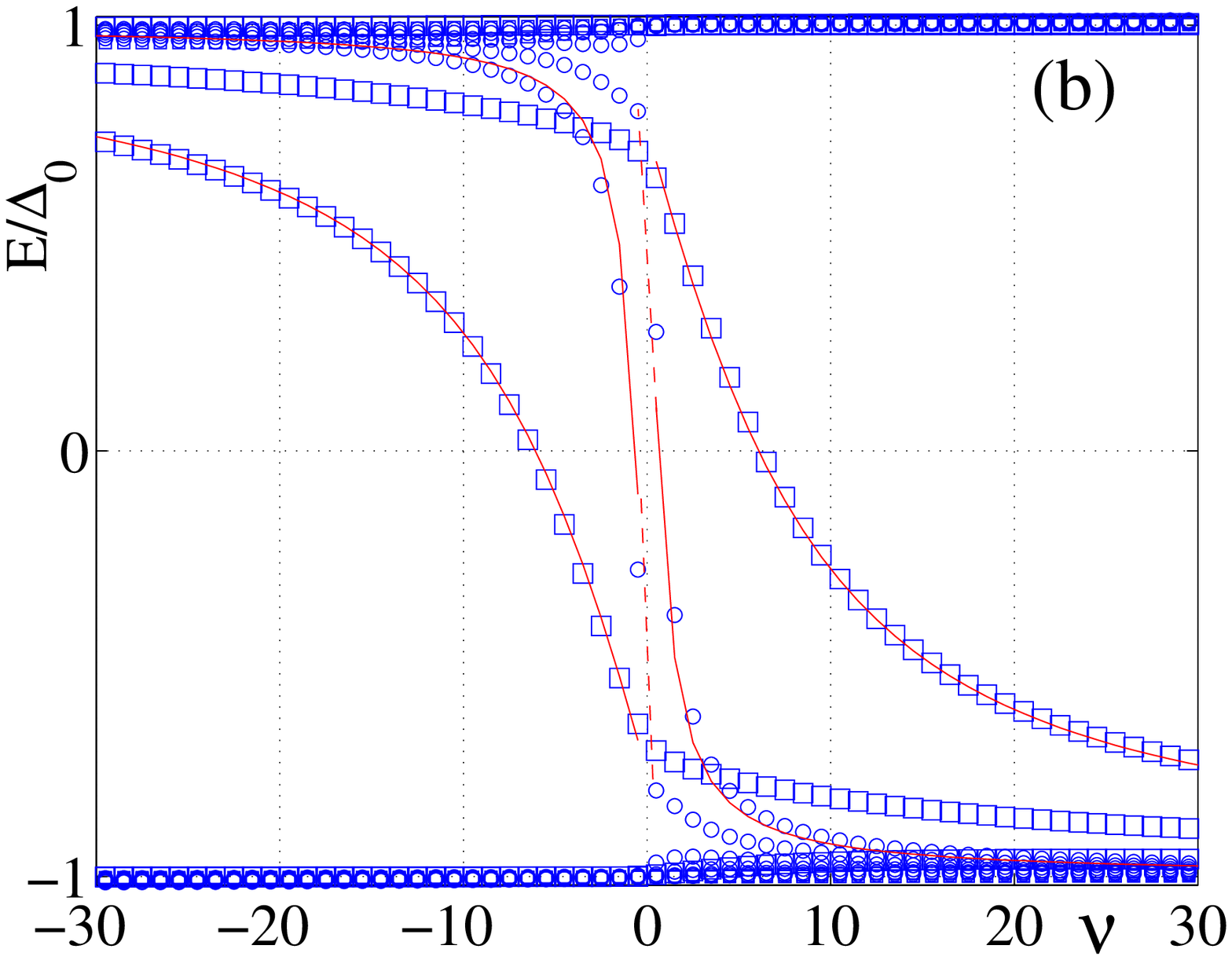}}
\caption{The subgap spectrum vs the angular momentum $\nu$ for a
singly (a) and doubly (b) quantized vortex in  superconducting
graphene with different doping levels: $\mu/\Delta_0=1$ (circles)
and $\mu/\Delta_0=10$ (squares). The quasiclassical CdGM solutions
of Eq.~(\ref{andreev}) are shown by solid lines (dashed parts of
lines are guides for eye). We choose here $R/\xi=100$. }
\label{Fig1}
\end{figure}
The normal--state eigenfunctions satisfying the above boundary
conditions are: $\hat U_{\nu, p}^e \propto \hat
U_{m_e,p}^{(0)}(\rho)$, $\hat V_{\nu, p}^e=0$ and $\hat U_{\nu,
p}^h=0 $, $\hat V_{\nu, p}^h\propto \hat U_{m_h,p}^{(0)}(\rho)$
where
$$
\hat U_{m,p}^{(0)}(\rho)=\left(\begin{array}{c} J_{m}(k_p^{m}\rho)
\\-J_{m+1}(k_p^{m}\rho)\end{array}\right)
$$
and $m$ is either $m_e$ or $m_h$, $m_{e,h}=\nu -(1\mp n)/2$, while
$k_p^{m}R$ are the $p$--th zeros of the Bessel function
$J_{m}(x)$. In a finite-size disk the basis also includes the
electronic and hole functions: $ \left( 0\atop \rho^{-m-1}\right)
$ for $m_{e,h}\leq -1$, respectively, which correspond to
$k_0^{m_{e,h}}=0$. The gap function of $n$-quantum vortex is
approximated by $ \Delta(\rho) = \Delta_0\, \mathrm{e}^{i n
\phi}[\rho  /\sqrt{\rho^2+\xi^2}]^n $. For the numerical
diagonalization procedure the matrix is truncated keeping the
number of eigenstates larger than $10 R/\xi$. Shown in Fig.~1 are
typical energy spectra for singly and doubly quantized vortices
calculated for different doping levels. Spectrum consists of
anomalous branches crossing zero of energy and branches lying
close to the gap edges. Figures 1(a) and 1(b) clearly demonstrate
the difference between odd and even vorticities. Fig.~1(a)
illustrates also a crossover to a spectrum of the CdGM type with
increasing doping level. For large $\mu$ the two anomalous
branches in a doubly quantized vortex are well described by the
quasiclassical analytical solutions found in Ref.\cite{MRS}.

\paragraph{Discussion.}
We can summarize that in an s-wave superconducting graphene the
subgap spectrum in a vortex core has a set of energy branches
which cross Fermi level as functions of $\nu$ treated as a
continuous parameter. The number of the branches is determined by
the vortex winding number $n$. Whether the crossing points of
these branches with zero belong to the energy states or not
depends on the parity of the vortex and on the doping level $\mu$.
For a degenerate case $\mu =0$ there are $|n|$ zero energy levels,
i.e., all crossing points belong to the spectrum. If the doping
level $\mu$ is increased, the spectrum transformation depends
strongly on the parity of the winding number. (i) For an odd
winding number one of the zero energy modes (type I mode), namely
that corresponding to $\nu = 0$, survives with an increase in the
doping level. The other $|n|-1$ (i.e., even number of) levels
split off zero with increasing $\mu$ (type II modes). (ii) For
even-quantum vortex, all $|n|$ levels belong to the second type
and split off zero. In general, the zero energy modes of the
second type can appear or disappear with the change in the Fermi
momentum and may exist only in pairs.

\acknowledgments

We are thankful to G.E. Volovik for many stimulating discussions.
This work was supported, in part, by Russian Foundation for Basic Research,
by the ``Dynasty'' Foundation and by the Academy of Finland (ASM).

\end{document}